\documentclass[twocolumn]{webofc}
\usepackage{enumitem}
\usepackage{color}
\usepackage{geometry}
\usepackage{enumitem}
\usepackage{comment}
\usepackage{braket}

\usepackage[varg]{txfonts}   
\begin{document}

\title{Ultra High Energy Cosmic Ray Source Models: Successes, Challenges and General Predictions}
%
%

\author{\firstname{No\'emie} \lastname{Globus}\inst{1}\fnsep\thanks{\email{noglobus@ucsc.edu}} \and
        \firstname{Roger} \lastname{Blandford}\inst{2}\fnsep\thanks{\email{rdb3@stanford.edu}}
}

\institute{Department of Astronomy and Astrophysics, University of California, Santa Cruz, CA 95064, USA
\and
           Kavli Institute for Particle Astrophysics and Cosmology (KIPAC), Stanford University, Stanford, CA 94305, USA
          }

\abstract{
Understanding the acceleration of Ultra High Energy Cosmic Rays is one of the great challenges of contemporary astrophysics. In this short review, we summarize the general observational constraints on their composition, spectrum and isotropy which indicate that nuclei heavier than single protons dominate their spectra above $\sim 5\,{\rm EeV}$, that they are strongly suppressed above energies $\sim50\,{\rm EeV}$,  and that the only significant departure from isotropy is a dipole. Constraints based upon photopion and photodisintegration losses allow their ranges and luminosity density to be estimated. Three general classes of source model are discussed  - magnetospheric models (including neutron stars and black holes), jet models (including Gamma Ray Bursts, Active Galactic Nuclei and Tidal Disruption Events) and Diffusive Shock Acceleration models (involving large accretion shocks around rich clusters of galaxies). The value of constructing larger and more capable arrays to measure individual masses at the highest energies and probably identifying their sources is emphasized. 

}
\maketitle

\section{Introduction}
\label{intro}

Understanding the acceleration of atomic nuclei to macroscopic energies $\gtrsim50\,{\rm J}$ is one of the great challenges of modern astrophysics. These Ultra High Energy Cosmic Ray (UHECR) accelerators operate undercover because the Universe is filled with electromagnetic and hydromagnetic waves that scatter and attenuate the cosmic radiation making direct source identification nearly impossible. On the other hand, these interactions lie at the heart of cosmic-ray science because they are intimately involved in most acceleration models. They also lead to the production of secondary messengers -- $\gamma$-rays and neutrinos -- signatures of leptonic and hadronic processes at play in  cosmic accelerators. The modeling of these interaction processes connects particle physics to multimessenger astronomy which can reveal the UHECR sources indirectly.
$\gamma$-rays and X-rays can be produced by both hadronic and leptonic processes, and a  multimessenger effort is made to detect  high energy neutrinos as messengers of hadronic accelerators.

In this brief review, we summarize direct and general observational constraints on UHECR sources (Sec.~2). In Sec.~3, we discuss three general UHECR acceleration sites: magnetospheres, jets and non-relativistic shocks. Magnetospheres are mostly associated with compact objects -- spinning neutron stars and black holes -- and can plausibly accelerate cosmic rays to the observed energies. The challenge is to explain how they can escape without loss. Jet models for UHECR acceleration are mostly associated with black holes jets associated with Gamma Ray Bursts (GRB), and Active Galactic Nuclei (AGN). Here the challenge is to identify sources within the UHECR horizon, as these sources are rare. Shock models are most challenged by the need to accelerate  cosmic rays to the highest energies measured. A hierarchical approach in which the entire observed spectrum originates successively at SuperNova Remnant (SNR) shocks, Galactic Wind Termination Shocks (GWTS), and at Cluster Accretion Shocks (CAS), is outlined. The opportunity for deciding between these, other, possible acceleration sites in the near future is discussed in Sec.~4.

\section{Summary of Observational Constraints on the origin of UHECR}
Using observations made with the Telescope Array (TA)~\citep{2008NuPhS.175..221K}, and the Pierre Auger Observatory (PAO)~\citep{PierreAuger:2004naf}, over the past decades, it has been established that nuclei are dominating the spectrum already at energies  $\sim5$~EeV. Both experiments confirm the existence of an "ankle" in the cosmic-ray spectrum at $\sim$~5 EeV. Both experiments agree that the observed spectrum starts to be strongly suppressed at energy $\gtrsim50\,\,{\rm EeV}$. This could be due to photopion production and photodisintegration by microwave background photons --- the GZK effect \citep{Greisen1966jv,Zatsepin1966jv} --- or may represent the maximum energy that can be accelerated in any cosmic source.

Despite many claims of association of UHECR with specific sources or source classes, the only significant departure from complete isotropy of UHECR is a $6.6\,{\sigma}$ dipole reported by PAO at energies above 8~EeV\citep{2017Sci...357.1266P}. 
It has been shown that the evolution of the dipole amplitude with energy is expected if the sources are extragalactic, as a result of the evolution of the GZK horizon with energy~\citep{Globus:2017fym}.
This feature  could reflect an anisotropy in the distribution of many sources following the large scale structure \citep{2021ApJ...913L..13D}, local sources such as nearby radio galaxies \citep{Eichmann_2022} or the presence of a single dominant source, such as a Galactic transient~\citep{Eichler:2016mut} -- which may require strong  Galactic magnetic turbulence to not upset the anisotropy limits but cannot be totally ruled out. However, the source location could well be significantly shifted across the sky due to cosmic ray deflection  by the poorly understood strong Galactic and extragalactic magnetic fields.  Given the uncertainty between the magnetic field models, one cannot reach strong conclusions yet regarding the nature of the sources with the dipole anisotropy \citep{Allard:2021ioh}.

It should be noted that while the dipole appears above 8~EeV, the sky in the 4-8 EeV energy range is compatible with isotropy \citep{2017Sci...357.1266P}.  This is certainly  due to the  combined effect of a large GZK horizon and the presence of accelerators operating up to lower maximum rigidity. The ``light ankle'' reported by the KASCADE Grande experiment at $\sim0.1$~EeV \citep{Schoo:2016yeo}  may either be a distinct proton component from more distant sources, or the accumulation of secondaries from the primary UHECR accelerators \citep{PhysRevD.92.021302}.

Another significant and generally accepted feature of the observations is the change in the cosmic ray composition with energy (see Figs~2.10-2.12~in \citep{Coleman_2023}). Such a statement can only be made on statistical grounds because we cannot determine the composition of individual primary cosmic rays. The nuclei become progressively heavier above the  ankle at $\sim5\,{\rm EeV}$. The cosmic rays are mostly heavy nuclei (CNO or heavier) at energies $\gtrsim40\,{\rm EeV}$ \citep{Coleman_2023}. There are too few showers with larger primary energy to be definitive, but it is generally supposed that the composition remains heavy up to the highest energies. 

This change in mindset by the UHECR community has a simple interpretation. Most source and propagation models depend upon the particle's rigidity, $R$, its momentum per unit charge, not its total energy. If we use rigidity instead of energy, the observed cosmic ray spectrum can be accounted for with a maximum rigidity at the accelerator $R_{\rm max}\sim8$~EV. To account for the whole extragalactic spectrum with only a single extragalactic component, one needs a very hard spectrum at the sources $R^{-1}$  to reproduce the evolution of the composition above the ankle, and a soft proton component ($\sim R^{-2.5}$ for sources avolving as the SFR \citep{globus2017b}) to account for the light ankle seen by KASCADE-Grande. Such a single-component model gives a coherent picture of the Galactic-to-extragalactic cosmic ray transition \citep{globus2015} (in this model the protons are the product of photodissociation processes at the source). Other models based on more than a single extragalactic component remain of course possible, and the sub-ankle protons could be the result of another type of accelerator operating in this energy range.
The contemporary luminosity density for UHECR acceleration is ${\cal L}_{\rm UHECR}\sim6\times10^{44}$~erg~Mpc$^{-3}$~yr$^{-1}$ above 5~EeV \citep{PhysRevD.102.062005}. For the purposes of astronomical comparison, this is very roughly $3\sim10^{-5}$ times the local galaxy luminosity density. More quantitative comparisons require specific source and propagation models.   

Regarding the cosmogenic gamma-ray diffuse flux (the by-product of the GZK effect), the mixed-composition model appear to be less constrained by the Fermi-LAT than the electron-positron dip (pure proton) scenario \citep{2011PhLB..695...13B, Gavish_2016, PhysRevD.94.063002} 
that rules out SFR-like and stronger cosmological evolutions. For a mixed composition, however, only very strong evolutions (e.g., FR-II radio galaxies) are excluded by the current observations \citep{2017ApJ}. The cosmogenic neutrino flux is well below the IceCube sensitivity for a mixed composition \citep{2017ApJ}. The IceCube collaboration reported a correlation between a PeV neutrino and a blazar \citep{IceCube:2018cha} implying that blazars are able to accelerate protons to $\sim 0.1$~EeV.

In summary, it is reasonable to assume that any source model needs to be able to accelerate cosmic rays up to a minimum rigidity $\sim$8~EV (equivalently iron at $\sim$200~EeV) to account for the observed spectrum and composition; therefore, in order to be considered as a possible source candidate, several conditions must be fulfilled by a cosmic accelerator:\\
${\small \bullet}$ Confinement:  the acceleration time should be smaller than the escape time; this is related to both the strength of the magnetic field and the geometry, size and age of the accelerator -- this is an expression of the famous Hillas criterion~\citep{1984ARA&A..22..425H};\\
${\small \bullet}$ Power: the central engine must be able to provide the necessary energy (either electromagnetic or kinetic, depending how the energy transfer is mediated);\\
${\small \bullet}$ Radiation losses at the accelerator: within the accelerating field the acceleration time should be smaller than the radiation loss time;\\
${\small \bullet}$ Interaction losses (photopion and photodisintegration) at the accelerator:  the acceleration time should be smaller than the interaction loss time;\\
${\small \bullet}$ Emissivity: the  density of sources must be enough to account for the observed UHECR flux (${\cal L}_{\rm UHECR}\sim6\times10^{44}$~erg~Mpc$^{-3}$~yr$^{-1}$ above 5~EeV) \citep{PhysRevD.102.062005};\\
${\small \bullet}$  Secondary radiation: the accompanying photon and neutrino flux should not be greater than the observed fluxes. This constraint must be satisfied by the secondary gamma-ray and neutrino flux at the sources and also by the cosmogenic, diffuse flux.  \\
${\small \bullet}$ Anisotropy: the source model needs to account for the observed anisotropies. This is the weakest constraint due to the uncertainty regarding the magnetic fields~\citep{Unger:2017kfh, refId0}.

\section{UHECR Source Models}\label{sec-source}
\subsection{Magnetospheric Models}
\subsubsection {General Considerations}
In order to accelerate a proton to an energy $\sim8\,{\rm EeV}$, it must  follow a trajectory where $\int d{\bf r}\cdot{\bf E}$ exceeds $\sim8$~EV. The simplest way to imagine this happening is in the context of a large scale magnetic field. Faraday's law guarantees that if the magnetic field varies with time, then there will be an EMF and an opportunity for particle acceleration. However, even a stationary flow can produce large, accelerating, electric field when there are appropriate boundary conditions. One such configuration is the unipolar inductor \citep{Weber}. A spinning, conducting body is envisaged to be endowed with an axisymmetric magnetic field distribution. In the simplest case, the body rotates with angular velocity~$\Omega$. 

If sufficient plasma is continuously produced to sustain an electrical current, without having dynamical relevance, the electromagnetic field in the ``magnetosphere'' will be force-free. Imposing a boundary condition at the neutron star surface leads to an electric field ${\bf E}=-({\bf\Omega}\times{\bf r})\times{\bf B}$ which implies that ${\bf E}\cdot{\bf B}=0$. The magnetic flux surfaces will be equipotential and isorotational and the potential difference between two adjacent flux surfaces containing magnetic flux $d\Phi$ is $dV=\Omega d\Phi/2\pi$. There will also be an electrical current $I(\Phi)$ flowing within a flux surface labeled by $\Phi$ which is associated with a toroidal component of magnetic field and an outward Poynting flow of electromagnetic energy, $IdV$. This Poynting flux extracts rotational energy efficiently from the neutron star with very little dissipation in the neutron star. It can be radiated at great distance, for example in a Pulsar Wind Nebula such as the Crab Nebula.  

The relationship between $I$ and $V$ depends upon the detailed boundary conditions but, if the electromagnetic field remains force-free, then $V\sim I Z_0/2\pi$, where $Z_0\equiv377\,{\rm Ohm}$ is the impedance of free space. To order of magnitude, a potential difference of, say, $10\,{\rm EV}$ is then associated with an electromagnetic power $\sim10^{36}\,{\rm W}$. Whether or not this potential difference is used for efficient UHECR particle acceleration depends upon circumstance but, if it is, then the associated power is inescapable.

\subsubsection{Neutron Stars}
Now consider the application of these general ideas to neutron stars. Spinning magnetized neutron stars are, manifestly, not axisymmetric but quite similar considerations apply. If the magnetic moment is that of a regular pulsar, then the spin period must be close to the maximal allowed value $\sim1.5\,{\rm ms}$ and the source lifetime is only years \citep{Fang:2012rx}. Magnetars, with surface magnetic field $\sim100\,{\rm GT}$ are more promising \citep{Arons:2002yj}. A spin period $\sim30\,{\rm ms}$ suffices but the source lifetime would be less than a day. It is hard to see how UHECR could escape the environs of a neutron star so soon after its birth given the high density of matter and radiation which should cause catastrophic losses. 

However, there is another possibility and this is that the power derives from the magnetic, not the rotational energy. This can be as much as $10^{41}\,{\rm J}$ and it could be released intermittently, following magnetic flares, where $\Omega$ would be replaced by the reciprocal of the time it would take light to cross a fraction of the magnetic surface, a few $\mu s$. Furthermore, the magnetic energy stored below the surface might be hundreds of times larger than that within the magnetosphere and accessible for much longer. This is important because the magnetar birth rate should be less than $\sim10^{-6}\,{\rm Mpc}^{-3}\,{\rm yr}^{-1}$. requiring an energy per magnetar of $>10^{43}\,{\rm J}$ to account for the UHECR luminosity density.

Despite this energetic discouragement, it is instructive to consider the general principles of particle acceleration exhibited by an axisymmetric pulsar magnetosphere. In this case, the magnetic field is stationary and the EMF formally vanishes. The magnetic surfaces are electrostatically equipotential. However, a current must flow which may require the continuous formation of electron-positron pairs in the magnetosphere. It has long been supposed that this can happen at a ``gap'', a possibly transient potential drop along the magnetic field where ${\bf E}\cdot{\bf B}\ne0$ and single charged particles can be accelerated to sufficient energy to produce $\gamma$-rays which can form pairs and breakdown the vacuum. There has been much modeling of gaps and generally it has been found that the associated potential drops across a gap are order of magnitude smaller than what would be needed to create UHECR directly at gaps. They could, of course be associated with the emission of much less energetic $\gamma$-rays and cosmic rays. 

If there is minimal potential drop along the flux surfaces, then the power must associated with electrical current crossing between the flux surfaces. There are many different ways that this has been modeled. For example, there could be enough plasma present that there is a Lorentz force density, ${\bf j}\times{\bf B}$, that gradually converts the electromagnetic energy into kinetic energy flux with minimal dissipation. This outflow might then be an efficient particle accelerator in a jet or wind (see below). Alternatively, the current may complete dissipatively as particle acceleration. At one extreme, all of the power may heat the thermal plasma and cause it to radiate. An intermediate possibility is that much of this power creates a suprathermal distribution of intermediate energy cosmic rays. 

A third possibility is that it is the highest energy particles that carry most of the electrical current across magnetic surfaces. The manner in which this may happen is interesting. A modest energy particle will gyrate around and move along the magnetic field in the local frame in which the electric field vanishes. However, for larger particle energies the gyro radius, $r_L$, increases and the collisionless particle motions also include drift velocities of their guiding centers with speeds $\sim cr_L/L$, where $L$ is the scale length. It is possible that despite the rarity of the highest energy particles, their greater mobility can allow them to carry most of the electrical current and, consequently, be accelerated to energies as large as those for which $r_L$ approaches $L$, specifically, those with energy $\sim ZecBL$. This basic mechanism has been invoked to account for intermediate energy particle acceleration at a pulsar wind termination shock. For this to be possible, in principle, requires that the electrical current flows along and not across magnetic surfaces all the way from the pulsar magnetosphere to the surrounding nebula. 

\subsubsection{Black Holes}
Many of these basic physics ideas are translatable to black hole magnetospheres with some important differences. We suppose that magnetic flux is trapped by the inertia and pressure of orbiting gas and some of this flux passes in and out of the event horizon. The black hole is not a perfect conductor. In fact, a formal surface conductivity can be associated with the horizon associating a resistance of $\sim Z_0/2\pi$ with the black hole. This has two consequences. The magnetic field lines rotate with an angular velocity roughly half that of the black hole. There is as much power dissipated behind the horizon as there is extracted outside the black hole. 

For the case of massive black holes, the power associated with UHECR acceleration is then, at least, comparable with that radiated by a moderate AGN such as a FRI radio source or a Seyfert galaxy. There is the same need to supply electrical charge continuously and gaps have been invoked as sources of $\gamma$-rays (e.g. \citep{Levinson:2010fc}). As with pulsars, the potential differences that need to be applied to keep the current flowing are generally tiny compared with those needed to account for UHECR. To drive home this point, if the current carrying pair plasma had the minimum density and energy $\sim1\,{\rm MeV}$ required, then its energy density would be only a fraction $\sim$ the ratio of this energy to the total potential difference --- $\sim1\,{\rm MeV}$ to $\sim10\,{\rm EV}$ or $10^{-13}$. (This is also the ratio of the minimum to the maximum gyro radii.) In fact gaps will not even be required if either the magnetic field lines are able to undergo interchange instability to keep the magnetosphere supplied with electron-ion plasma or externally produced $\gamma$-rays can produce pairs directly. For all of these reasons, gaps are not seen as UHECR sources. 

As with pulsars, it also possible to imagine particle acceleration occurring when current crosses the toroidal magnetic field lines associated with the electromagnetic outflow produced by a magnetized, spinning black hole. This takes place in the ``Espresso'' model of UHECR acceleration \citep{Caprioli_2015} where lower energy cosmic rays are injected into the outflow and emerge with rigidity comparable with the total potential difference available --- an expression of the famous Hillas criterion \citep{1984ARA&A..22..425H} (see section 3.2.1 below).  Assuming $\Gamma\sim30$ (as in powerful blazars), a one-shot boost of a factor of $\sim \Gamma^2$ in energy, can transform the highest-energy galactic CRs at 100 PeV in the highest-energy UHECRs at 100 EeV. If this mechamism can account for the whole UHECR data is still an open question.

There is a second, quite general constraint on locating UHECR acceleration close to an AGN black hole. A $\sim100\,\mu$ far infrared photon will be able to create pions in the Coulomb field of a $\sim10\,{\rm EeV}$ cosmic ray. The UHECR acceleration site must be sufficiently far away from the AGN that the photon density does not prevent energy loss or photodisintegration. Similar remarks apply to the pair production threshold which involves $\sim30\,{\rm GHz}$ radio photons. As discussed below, even more stringent constraints must be satisfied to avoid photodisintegration of the heaviest nuclei. The end result is that the only way that processes directly connected to an AGN magnetosphere could account for UHECR is if they involve low luminosity AGN and operate at some remove from the actual black hole magnetosphere.

Similar arguments can be used to rule out UHECR production around or directly associated with the magnetospheres of stellar mass black holes either close to formation in GRB or later on in binary X-ray sources. We now turn to the viability of UHECR acceleration further away from massive or stellar black holes, within relativistic jets. 

\subsection{Relativistic Plasma Jet Models}

Relativistic jets are ubiquitous in the universe and their acceleration, collimation  and deceleration processes provide  a variety of  acceleration sites for UHECRs. The central engine of relativistic jets  is still under debate. Two driving mechanisms are commonly invoked,  heating mediated by accretion  ("nurture"), which  leads to  pressure-driven jets, and the Blandford-Znajek mechanism  where the jet is powered by a rapidly spinning black hole ("nature") which leads to magnetically-driven  jets \citep{Blandford:2022pwc}. The magnetization factor has important implications for modeling the acceleration of UHECRs and the production of secondary $\gamma$-rays and neutrinos. 

Because the jets are structured both laterally but also temporally, the  ejected  plasma layers have different velocities and hence shear acceleration or diffusive shock acceleration can take place at internal shocks in the jets, lobes, or at the oblique shocks in the jet's boundary layer, or far beyond the Bondi radius at  external shocks when the jet decelerates in the ambient medium. These mechanisms allow the transfer of the kinetic energy of the relativistic outflow to  cosmic rays. This is mediated by magnetic turbulence and the micro-physics at play in generating the magnetohydrodynamic waves  is important and yet poorly constraint in relativistic jets.  

Efficient dissipation of magnetic energy can happen through turbulence or reconnection and leads to the emission of high energy photons that can interact with the cosmic rays via photointeractions (pion production for protons or Giant Dipole Resonance, GDR, for nuclei).  How this operates  in relativistic outflows is still unsettled and depends on the configuration of the magnetic fields. The two configurations commonly invoked are: a large scale, ordered field of a single polarity, leading to a highly magnetized relativistic jet;  or a magnetic field of  alternating polarity leading to a "striped" jet made of regions with toroidal field of alternating polarity. 
Instabilities such as the kink can also drive magnetic dissipation. Therefore, for relativistic jets, the following questions remain: What is the best sites for UHECR acceleration in relativistic jets?  Are the heavy nuclei able to survive to the high energy photon fields present around relativistic sources? Let us review below some of the most promising jetted source candidates for UHECR sources.

\subsubsection{Jet Power requirement for UHECR}
Let us recall the jet power requirement for UHECR production. In the following we consider a relativistic plasma jet with half opening angle is $\theta$, $r$ is the jet height above the launching region. We use masses in units of solar mass $\bar{m}={M}/{M_\odot}$, luminosities in units of $L_{Edd}=1.26\,10^{38}\bar{m}$ erg/s and radii in units of gravitational radius $\bar{r}={r}/{r_g}$ with $r_g=3\cdot10^{5} \bar{m} $ cm. 
Quantities in the jet comoving frame are  denoted with a prime symbol.
The jet expansion time in the comoving frame is $t'_{\rm exp} = r/(\Gamma\beta c)$, where $\Gamma$ is the bulk Lorentz factor of the jet $\Gamma=(1-\beta^2)^{-1/2}$.
The condition for jet stability implies that the comoving signal crossing timescale is smaller than $t_{\rm exp}$, 
which gives the condition $\Gamma\theta\lesssim\beta_s/\beta$ where $\beta_s c$ is the Alfv\'en velocity. In relativistic magnetized jets $\beta\approx\beta_s\approx1$. The acceleration region cannot extend further than the transverse distance $r\theta$. The escape time sideways is thus $t'_{\rm esc} = (r\theta)/(\Gamma c)$. 

The jet luminosity $L_{j}=\pi(\theta r)^2\beta\Gamma^2cu'_j$ where $u'_j$ is the comoving energy density.
The magnetic field of a relativistic jet with magnetic luminosity $L_B\equiv\xi_B L_{j}$, bulk Lorentz factor $\Gamma$ and half opening angle $\theta$ is
\begin{eqnarray}
B'&=&\frac{2\left(\xi_B L_j/c\right)^{1/2}}{\theta\Gamma r}\nonumber\\
&\approx& 4.3 \cdot 10^8 (\xi_B\bar{L}_j/\bar{m})^{1/2}\bar{r}^{-1}(\theta\Gamma)^{-1}{\rm G}\,.
\label{Bfield}
\end{eqnarray}
The magnetic field $B'$ is related to the total energy density $u'_{j}$ by: 
$B'=\sqrt{4\pi\xi_Bu'_{j}}$.

The acceleration time  $t'_{acc}= t'_L/\xi_{acc}= E'/(ZeB'c\xi_{acc})$ with $\xi_{acc}\lesssim1$ (in numerical experiences $\xi_{acc}\lesssim0.1$ making the usually assumed Bohm diffusion overoptimistic).  
The condition $t'_{acc}=t'_{esc}$ (following \citep{1984ARA&A..22..425H}) gives:
\begin{eqnarray}
E'_{\rm max}&=&\frac{\xi_{acc}r\theta Z e B'}{\Gamma}\approx300\,  \xi_{acc}Z[B'_{\rm G}r_{\rm cm}](\Gamma\theta)\Gamma^{-2}\, {\rm eV}\nonumber\\&\approx& 3.\cdot10^{16}  \xi_{acc} Z\left(\xi_B\bar{L}_j\bar{m}\right)^{1/2}\Gamma^{-2} {\rm eV}\,,
\label{emax_conf}
\end{eqnarray}
which gives the condition on the observed jet luminosity:
\begin{equation}
L_{j}>8.4\cdot10^{44}\xi_B^{-1} \Gamma^{2}\xi_{acc}^{-2}\left(\frac{E_{\rm max}}{Z\,10^{20}{\rm eV}}\right)^2 {\rm erg \,s^{-1}} \ , 
\label{emax_conf}
\end{equation}
where $E_{\rm max}/Z$ is the maximum rigidity reachable in the observer's frame.

{ It is important to note that here only the mean intensity of the magnetic field $B'$ is considered.
However in the usual model of diffusive shock  acceleration (DSA), the largest turbulence scale of the magnetic field plays  an important role, as the relevant escape time is the escape time {\it upstream} of the shock front.
As the shocks microphysics is poorly understood (especially in the mildly-relativistic regime), the usual prescription for the particles confinement is that a cosmic ray remains accelerated at the shock if the Larmor radius of the particle is smaller than the largest turbulence scale of the magnetic field in the shocked region, i.e. $r_L\lesssim \lambda_{\rm max}$.
It is important to keep in mind that $\lambda_{\rm max}$ can be a small fraction of the size of the system. Therefore, in the case of DSA, $t'_{\rm esc} = \xi_{esc}(r\theta)/(\Gamma c)$ with $\xi_{esc}\lesssim1$.\\
Also, the above estimate of the maximum rigidity reachable is based on a geometrical argument. However there are several energy losses that will limit the acceleration process.
The maximum comoving energy $E'_{\rm max}$ of the accelerated particles is estimated by equating their acceleration time $t'_{acc}$ with the relevant loss time $t'_{loss}$, and the related "opacity" is defined as $\tau_{loss}\equiv{t'_{esc}}/{t'_{loss}}$. 

 }
The synchrotron opacity is given by
\begin{equation}
\tau_{syn}=\frac{(r\theta)Z^4\sigma_TB^2E}{6\pi m_e^2c^4\Gamma}\left(\frac{m_e}{m_A}\right)^4\,.
\end{equation}

Concerning photodissociation, the lowest energy and highest cross-section process is the giant dipole resonance.
The energy loss rate due to GDR  interaction with the photons of energy $\epsilon'$ is 
$(t'_{\rm GDR})^{-1}\simeq\sigma_{GDR}n'(\epsilon')c{\Delta\epsilon_{\rm GDR}}/ {\epsilon_{\rm GDR}} \,\,{\rm s^{-1}}$
with $\sigma_{GDR}=1.45\,10^{-27}A$ cm$^2$, $\Delta\epsilon_{\rm GDR}=8$ MeV and $\epsilon_{\rm GDR}\approx 42.65A^{-0.21}$ MeV.
The photon energy $\epsilon'$ is given by the threshold condition for GDR interaction: $E'_{\rm N}(A)\epsilon'\approx2.5\,10^{17} (A/56)$ eV$^2$. 
The GDR opacity is $\tau_{GDR}\equiv{t'_{esc}}/{t'_{GDR}}$. 
With an opacity  close to 1, we are in the optimum working condition for the energy loss process considered. Only when the shock becomes transparent the maximum rigidity is limited by the escape from the magnetized region. 
Particle escape acts as a high pass filter: only particles close to the maximum rigidity $\simeq E'_{\rm max} /Z$ can escape from the magnetized region upstream of the shock wave. We expect that the spectrum for the escaping particles is thus very hard, almost a delta function around $\simeq E'_{\rm max} /Z$.

\subsubsection{Gamma-Ray Bursts Jets}

Gamma-ray bursts (GRBs) are associated with relativistic jets from newly born stellar-mass black holes. There are three different types of GRBs, which are  associated with different progenitors. Long GRBs (lGRBs) are associated with the core-collapse of massive stars (more precisely, to SN Ib/Ic). They are the most powerful type of GRBs, with a jet luminosity able to reach $\sim 10^{52}$~erg/s. They are also the rarest, with a rate of $\sim 1$~Gpc$^{-3}$yr$^{-1}$ and a $\gamma$-ray luminosity density $\sim 6\, 10^{42}$~erg Mpc$^{-3}$yr$^{-1}$. Short GRBs (sGRBs) are the aftermath of binary neutron star and black hole neutron star mergers and have jet luminosities of typically $\lesssim 10^{49}$~erg~s$^{-1}$. Finally, there is another class of GRBs, the low luminosity GRBs (LLGRBs) which have lower luminosities (typically $L_{\rm jet}\lesssim 10^{50}$~erg~s$^{-1}$,  $\sim100$ times the rate of lGRBs \citep{2007ApJ...662.1111L}.

In the case of GRBs jets, there are two types of shocks that have been widely considered for diffusive shock acceleration: external shocks and internal shocks \citep{1995PhRvL..75..386W,1995ApJ...453..883V}. External shocks are caused by the jet deceleration in the ambient medium (these are responsible for the GRB afterglow) and are superluminal.   Many authors pointed out the inefficiency of Fermi acceleration at ultra-relativistic shocks, unless a strong small-scale turbulence is present in the downstream medium \citep{2006ApJ...641..984N,2006ApJ...645L.129L}.  Acceleration in ultra-relativistic external shocks   can produce cosmic rays  with energies above a few $Z\times 10^{15}$ eV (PeVatrons) 
see e.g., \citep{1999MNRAS.305L...6G,2021ApJ...915L...4G}- but not the highest energy cosmic rays.

Internal shocks are expected to form inside the jet once the fast layers of the plasma jet catch up with the slower parts. In that case, the shock in the comoving frame is mildly relativistic (with Lorentz factor of $\sim$1.1 to $\sim$2) so efficient diffusive shock acceleration can take place. It has been shown that the spectrum of UHECR emitted by GRB internal shocks can reach maximum rigidities of 8~EV (iron at $\sim$100 EeV) for the highest luminosities ($L_{\rm jet}\gtrsim 10^{52}$~erg/s), making diffusive shocks acceleration at internal shocks a promising candidate for UHECR production \citep{10.1093/mnras/stv893}. However, the caveat for the long GRB model is that they fall short to account for the required UHECR luminosity density (typically by a factor $\sim100$) unless they have a poor $\gamma$-ray efficiency \citep{2010ApJ...722..543E,10.1093/mnras/stv893}; or, it has been proposed that one single Galactic event can be responsible for most of the UHECR flux if the local extragalactic magnetic field is large and can trap efficiently $\lesssim$8 EV rigidity particles \citep{Eichler:2016mut}.

\subsubsection{Active Galactic Nucleus Jets}
The jets associated with massive black holes in active galactic nuclei, where the outflow is magnetically independent of a central black hole/accretion disk source, have also been proposed as UHECR accelerators. (By magnetically independent, we mean that the large-scale electrical current that flows along the inner jet has mostly left the jet, to be replaced by small scale current supporting small scale magnetic field.) Here two main mechanisms have been proposed, relativistic shock fonts and magnetic reconnection in a relativistic boundary layer. As just pointed out, relativistic shocks are qualitatively different from their nonrelativistic counterparts and they may be much less efficient in accelerating the highest energy particles. Partly for this reason, it has been suggested that it is the nonrelativistic backflow after a relativistic jet passes through a relativistic termination shock that is where more efficient UHECR acceleration can occur \citep{10.1093/mnras/sty2936}. As the number of suitable sources within the UHECR horizon is small, there is the future possibility of making tentative associations with the highest energy particles may be possible \citep{Bell:2021pkk}.

There is good observational evidence that relativistic jets decelerate, dissipate, accelerate and radiate through their surfaces, especially close to their massive black hole sources. Presumably gas is entrained into the electromagnetic jet from a much more slowly moving external medium. These can be characterized as boundary layers from a gas dynamical perspective and as a cylindrical, relativistic current sheet from an electromagnetic perspective. It is presumably the second description that is most promising as an accelerator. Either way, this is a site where magnetic reconnection will take place. Non-relativistic reconnection is a process which has been extensively studies and there is much space physics data to compare with simulations. Overall it is not very efficent in creating very high energy particles. However, relativistic acceleration has recently been studied quite extensively and it involves novel features, especially the launching of high speed plasmoids \citep{Sironi:2020mzd}. There does not seem to be any natural way to accelerate particles to the very highest energy but this is certainly worth further consideration.

\subsubsection{Tidal Disruption Events Jets}
There has been much recent observational attention paid to Tidal Disruption Events (TDE), where individual stars orbit massive ($M\lesssim10^8\,{\rm M}_\odot$) black holes in otherwise dormant galaxies passing by so close that they are they are ripped apart by tidal forces with the debris falling back onto the black hole on timescales of months to years and radiating in the optical and X-ray bands especially. A minority of TDEs are found to be jetted and there is evidence that the jets have Lorentz factors $\gtrsim10$, like blazars \citep{10.1093/mnras/stad344}. They have also been proposed as sources of UHECR \citep{Farrar:2014yla}.  

The estimated rate of jetted TDEs beamed towards us is $2\times10^{-11}\,{\rm Mpc}^{-3}\,{\rm yr}^{-1}$ \citep{Andreoni:2022afu}. A conventional estimate of the total rate of jetted TDE is perhaps a hundred times larger. On this basis, we can estimate the energy required per jetted TDE to account for ${\cal L}_{\rm UHECR}$ (Sec.~2) as $\sim0.03\,{\rm M}_\odot c^2$. This is not impossible but it seems implausible given the very much lower radiative efficiencies typically inferred.

\bigskip

\subsection{Hierarchical Acceleration Model}
\subsubsection{Diffusive Shock Acceleration}
An alternative class of UHECR acceleration models invokes Diffusive Shock Acceleration (DSA, see \citep{BE87} for a review) at giant intergalactic shock fronts, most notably those surrounding rich clusters of galaxies but also at shock fronts surrounding the filaments which connect these clusters. There are many possibilities. Here, we will strict attention to the more limited but more prescriptive and, consequently, more refutable idea that the entire cosmic ray spectrum observed at Earth is the result of successive DSA from distinct source classes. 

Specifically, we consider a hierarchical model in which the observed $\sim1\,{\rm GeV}-\sim3\,{\rm PeV}$ cosmic rays originate in nearby supernova remnants. The pressure from these cosmic rays drives a magnetocentrifugal Galactic wind which passes through a Galactic Wind Termination Shock (GWTS) where re-acceleration takes place. Some of the highest energy particles from the GWTS escape upstream and contributes to the shin part of the spectrum; the remainder are transmitted into the circumgalactic and intergalactic media. GWTS acceleration is even more powerful around galaxies associated with AGN and starbursts. 

These intermediate energy cosmic rays permeate the intergalactic medium and are input to DSA at the large intergalactic shocks. This ``holistic’’ interpretation features propagation as well as acceleration. Of course, there are many alternative acceleration sites that could dominate different parts of the spectrum in a hierarchical scheme, as discussed above, or, alternatively, the intergalactic acceleration might operate mainly on mildly relativistic particles injected directly at the shock front.  

DSA is a specific mechanism for converting a significant fraction of the kinetic energy flux, measured in the frame of the shock, of the gas upstream of a strong shock front into high energy particles that are transmitted downstream. It relies upon scattering by hydromagnetic disturbances/waves. Individual particles of rigidity (momentum per unit charge) $R$, with mean free paths $\ell(R)$, will diffuse against the flow with an upstream scale height $L\sim\ell c/u$, where $u$ is the speed of the shock relative to the upstream gas. In the simplest, test particle version of this mechanism at a planar shock front, relativistic particles will increase their rigidity by a fractional amount $\Delta R/R\sim4(1-1/r)u/3c$, where $r$ is the shock compression ratio, every time they make a double crossing of the shock front. On kinematic grounds, the probability that an individual cosmic ray not return upstream is $4u/rc$. The probability that a relativistic particle incident upon the shock front with initial rigidity $R$ be transmitted downstream with rigidity greater than $xR$ is 
\begin{equation}
P(>x)=x^{\frac3{r-1}};\quad x>\ge1.
\end{equation}
In this limit the shock behaves as a linear system, convolving the input upstream momentum space particle distribution function with a power law, Greens’ function to give the transmitted downstream distribution function. The slope of the power law does not depend upon the details of the scattering and no particles escape upstream. The maximum rigidity, $R_{\rm max}$ that can be accelerated by a given shock is usually determined by the condition that $L$ must be less than the size of the shock, typically its radius of curvature. Note that this condition is, at least, $c/u$ times more stringent that the Hillas criterion. Alternatively, it takes $\sim(c/u)^2$ scatterings, not one orbit, to double a particle's energy. DSA is inherently slower than the more violent mechanisms discussed above. The question is ``Can it still account for the highest energy particles we actually observe?''. 

Realistically, the test particle approximation fails in many ways. The acceleration is influenced by the cosmic-ray pressure gradient decelerating the gas ahead of the shock and changing the kinematics. In addition the cosmic rays can account for a significant fraction of the total energy of the flow and, on account of their different effective specific heat ratio, change the determination of $r$ through the conservation laws. The accelerated cosmic rays also create and sustain the wave turbulence that scatters them \citep{10.1093/mnras/172.3.557}. The most direct, and, arguably, the most important, form of scattering involves waves with wavelengths resonant with the particle gyro radii $r_{\rm L}(R)$. If the rms magnetic field amplitude of these waves is $\delta B(r_{\rm L})=\delta B(R)$, then $\ell\sim(B/\delta B)^2r_{\rm L}$, assuming random phases and Gaussian statistics (both questionable). If we assume that the waves are nonlinear over some range of $R$, then $\ell\sim r_{\rm L}$ and we call this Bohm diffusion. The efficiency of cosmic ray acceleration, $\epsilon_{\rm CR}$, which can be measured by the ratio of the downstream cosmic ray pressure to the upstream momentum flux, is then determined by the processes that determine the rms magnetic field strength. 

There are several possible sources of this wave turbulence. If cosmic rays stream with mean speed in excess of the Alfv\'en speed in a uniform magnetic field, then they will excite the growth of resonant Alfv\'en waves propagating along the direction of their mean velocity. These waves will then scatter the cosmic rays in pitch angle and reduce their mean motion so that it is not much more than the Alfv\'en speed. The linear growth rate is roughly $(n_{\rm CR}/n_i)c/r_{\rm L}$, where $n_{\rm CR}$ measures the number of resonant particles and $n_i$ is the background ion density. An alternative possibility is that waves are created hydromagnetically with wavelength much larger than $r_{\rm L}$ which initiates a turbulence spectrum where the waves with shorter wavelengths will have linear amplitudes. A third possibility is that the waves are generated with wavelengths shorter than $r_{\rm L}$ by the motion of electrons carrying the return current that must balance the current associated with the accelerated cosmic rays. Extensive ``PIC'' and related simulations have vindicated the generation of wave turbulence and elucidated how these and other processes interplay in DSA.

\subsubsection{Supernova Remnants}
Although we are mostly concerned with questions of UHECR acceleration, it is instructive to contrast intergalactic shocks with precursor acceleration of lower energy particles at smaller shock fronts. It is a reasonable conjecture that analogous physical processes operate at all high Mach number shocks, scaled by $L$ and the momentum flux. While there has long been strong evidence associating supernova remnants with the majority of Galactic cosmic rays up to PeV energy, the argument for the reminder of the spectrum being due to DSA, is no better than circumstantial. What is clear is that if DSA accounts for most of the total spectrum, the acceleration must occur with near maximal efficiency with respect to $R_{\rm max}$ and $\epsilon_{\rm CR}$ in all three sites. For this reason, it is interesting to look at this problem inductively --- to ask what would be required from the complex nonlinear plasma processes in the vicinity of the shock to account for the observations--- rather than deductively, to attempt to calculate these processes from first principles. 

Despite this commonality, there are important distinctions to be made. The first involves the kinematics. Most supernova remnants exhibit a quasi-spherical shock, convex on the upstream side, expanding into a near uniform interstellar medium, though stellar winds and molecular clouds can certainly introduce observable differences in many examples. This geometry facilitates the escape of the highest energy particle upstream as opposed to their transmission downstream, as assumed under the test particle approximation. Over half of the very highest energy particles can leave the remnant in this fashion. The original model of DSA essentially ignored these particles, supposing, instead, that the cosmic rays that we observed were transmitted downstream and were then released, following adiabatic decompression at the end of the remnant's lifetime. By contrast, it could be that most of the observed cosmic rays escape upstream with $R\sim R_{\rm max}$ which decreases as the remnant ages.

To be quantitative, the cosmic ray luminosity, per unit area, of the Galactic disk is generally estimated to be at least ten percent of the local supernova remnant luminosity per unit disk area. ($E_{\rm SNR}\sim10^{51}\,{\rm erg}$ released every $\sim30\,{\rm yr}$ throughout the Galaxy.) Now, in order to accelerate protons to $\sim3\,{\rm PeV}$ energy at a SNR expanding with speed $u_{\rm SNR}$ and radius $R_{\rm SNR}$, requires that the gyro radius be $\lesssim u_{\rm SNR}R_{\rm SNR}/c$ or $B\gtrsim100(u_{\rm SNR}/10,000\,{\rm km\,s}^{-1})^{-1}(R_{\rm SNR}/1\,{\rm pc})^{-1}\,\mu{\rm G}$, perhaps two orders of magnitude large than the rms Galactic field. Furthermore, the magnetic field strength is strongly bounded above by requiring that its contribution to the momentum flux, $B^2/4\pi$ be less than that of the gas, which is $\sim0.3E_{\rm SNR}R_{\rm SNR}^{-3}$, adopting the Sedov solution, or $B\lesssim10(R_{\rm SNR}/1{\rm pc})^{-3/2}\, {\rm mG}$. This interval allows efficient cosmic ray acceleration by young SNR up to the knee in the spectrum.

\subsubsection{Galactic Wind Termination Shocks}
The next level in this proposed hierarchy involves galactic winds. A large motivation for proposing a wind in our Galaxy comes from cosmic-ray observations. The measured ratio of light to medium nuclides at $\sim\,{\rm GeV}$ energy allows us to infer that these cosmic rays escape the disk after traversing a grammage $\sim6\,{\rm g\,cm}^{-2}$. This implies a residence time $\sim30-100\,{\rm Myr}$, much shorter than a disk rotation period. Cosmic rays leave the disk continuously and it is reasonable to suppose that they carry gas with them. The wind provides a chimney or exhaust for the cosmic rays (as well as the hot gas which is unable to cool radiatively) created by the SNR. The existence of an outflowing wind, occupying most of the volume of the Galactic halo, is not inconsistent with the presence of denser gas clouds falling inward, as observed.

Various models of this outflow have been entertained. One recent model posits the presence of a vertical magnetic field  at the disk and that $\sim\,{\rm GeV}$ cosmic rays stream along it just faster than the Alfv\'en speed so that Alfv\'en waves are excited and couple the cosmic rays to the gas, allowing it to achieve the escape speed from the solar radius. The cosmic rays alone drive the wind \citep{Mukhopadhyay:2023kel}. 

A more powerful wind is possible if the vertical magnetic field corotates with the disk. In this case the cosmic rays just elevate the gas to modest altitude $\sim$ a few kpc, at which point the field develops a large radial component and the gas is flung out magnetocentrifugally and it is the orbiting interstellar medium that provides most of the power for the wind. Under this scenario, the interstellar medium should be seen as an open system losing a significant fraction of its mass, and angular momentum as well as energy over a Galactic lifetime. This possibility is not generally entertained in models of the interstellar medium.

The outflow will pass through three magnetohydrodynamic (Alfv\'enic and magnetosonic) critical points before crossing a termination shock, typically at a radius $R_{\rm GWTS}\sim200\,{\rm kpc}$, with speed $u_{\rm GWTS}\sim1000\,{\rm km\,s}^{-1}$, before joining the circumgalactic medium. The flow expands with density $\rho$ decreasing by roughly three orders of magnitude. The cosmic ray rigidities decrease $\propto\rho^{1/3}$, roughly one order of magnitude. The poloidal component of the magnetic field decreases from $\sim2\,\mu{\rm G}$ at the disk to $B_p\sim3\,{\rm nG}$ at the shock, while the toroidal field is roughly ten times larger, $B_\phi\sim30\,{\rm nG}$.

The termination shock can reaccelerate $\sim{\rm GeV-PeV}$ cosmic rays to fill out the $\sim\,{\rm PeV-EeV}$ shin in the spectrum. The shock surface is likely stationary and concave on the upstream side with radius. This changes DSA in significant and observable ways. For modest cosmic-ray rigidity, say below the reduced peak of the cosmic-ray spectrum $R\lesssim3\,{\rm PV}$, the shock will be effectively planar and a reaccelerated spectrum will be transmitted downstream. However, when $\ell\gtrsim u_{\rm GWTS}R_{\rm GWTS}/c$, the high energy cosmic rays are able to cross the Galactic halo and continue their acceleration at a quite different part of the shock. This acceleration can continue up to a rigidity where $\ell\sim R_{\rm GWTS}$, perhaps $\sim100\,{\rm PV}$. When, as is usually the case, cosmic ray energy rather than rigidity spectra are displayed, the composition of the shin particles becomes progressively heavier, as is observed.

Again, to be quantitative, the wind may account for as much as $\sim1\,{\rm M}_\odot{\rm yr}^{-1}$mass loss from the galaxy and a corresponding power $\sim5\times10^{41}\,{\rm erg\,s}^{-1}$,  more than the cosmic ray power. A turbulent magnetic field strength $B_{\rm RMS}(100\,{\rm PV})\sim0.3B_\phi(R_{\rm GWTS})$ suffices to sustain acceleration within a closed shock surface throughout the shin observed at Earth. 

\subsubsection{Intergalactic Shock Fronts}
We now turn to the final and most relevant level in this simple hierarchy when the $\sim\,{\rm PeV-EeV}$ cosmic rays accelerated by galaxies, especially those that are more active than our Galaxy, are incident upon intergalactic shock fronts (e.g. \citep{1995ApJ...454...60N,1997MNRAS.286..257K}, Simeon et al. in preparation) . The most powerful shocks, prominent in cosmological simulations, surround rich clusters of galaxies which show Mach numbers that can exceed $\sim100$.  The assembly of clusters, over cosmological time, is complex but we will idealize the shock as a stationary spherical surface surrounding most of the cluster. These shocks have not been observed directly as yet. Deep searches at low radio frequencies seems well-motivated and would be very prescriptive, if successful. Simulations also show quasi-cylindrical filaments that may connect clusters and which may be surrounded by quasi-cylindrical infall shocks. The Milky Way may lie within such a filament.

The kinematics of a stationary cluster accretion shock, CAS, is different from that at SNR and GWTS. The convergence of gas ahead of the shock results in adiabatic heating of lower energy particles. This acceleration, which supplements DSA, will be less efficient for higher energy particles, which diffuse outward, relative to the gas flow. We continue to view the problem inductively and ask what scattering would have to be present in order for nearby cluster shocks to account for the observed UHECR spectrum.  If we take an accretion shock surrounding the Virgo cluster then we observation combined with cosmological simulations suggest a shock radius $R_{\rm CAS}\sim2\,{\rm Mpc}$ an infall speed and density ahead of the shock $u_-\sim1000\,{\rm km\,s}^{-1}$ and $\rho_-\sim10^{-29}\,{\rm g\,cm}^{-3}$. The shock is strong and the rate at which gas kinetic energy crosses it is $\sim10^{45}\,{\rm erg\,s}^{-1}$. If we adopt a rich cluster density of $\sim10^{-5}\,{\rm Mpc}^{-3}$, then the cluster gas kinetic energy luminosity density is $\sim30{\cal L}_{\rm UHECR}$. There seems to be sufficient power to account for the observed UHECR flux. The larger challenge is to accelerate heavy nuclei, optimally, to $R_{\rm max}\sim10\,{\rm EV}$.

In this model, we detect the UHECR that escape upstream with a spectrum peaking at a rigidity $\sim1-8\,{\rm EV}$ which will vary from cluster to cluster. The cosmic rays which are transmitted downstream will never be directly observable, though they might be seen at much lower energy through their $\gamma$-ray emission. In order to accelerate cosmic rays to $R_{\rm max}\sim10\,{\rm EV}$, we require that $R_{\rm CAS}>(c/3u_-)\ell(R_{\rm max}$. If we adopt Bohm diffusion, then  the rms, resonant field ahead of the shock on these scales is $B_{\rm rms}\sim1\,\mu{\rm G}$. This contrasts with a general field in the IGM that is generally expected to be $\lesssim1\,{\rm nG}$. The isotropic magnetic pressure associated with this field is roughly an order of magnitude smaller than the nominal momentum flux carried by the background gas $\sim10^{-13}\,{\rm dyne\,cm}^{-2}$. Provided that the turbulence does not dissipate and pre-heat the gas, thereby reducing the shock Mach number and the efficacy of DSA, a maximum rigidity $R_{\rm max}\sim10\,{\rm EV}$ seems attainable. However, if, for example, protons with significantly larger rigidity than this are identified, it seems hard to see how accretion shocks around relatively well-observed and simulated, local clusters of galaxies could account for their acceleration. 

Now turn to the input. Of course, we do not know what is the intergalactic spectrum produced by all the varied types of galaxy that surround the clusters. There are many possible contributors that we we may not be observing at Earth, for example starbursts and the winds they drive, pulsar wind nebulae, jets associated with spinning black holes and a network of weaker intergalactic shock fronts. (It is possible to associate the observed light ankle with filament shock acceleration in which case this comprises the spectrum transmitted downstream.) If we just confine our attention to the SNR spectrum produced by our local Galactic disk, then we expect a source spectrum $S(R)\propto R^{-2.2}$, as produced by all interstellar shocks over their lifetimes. This should extend up to a rigidity of order a few PV. At this point the rigidity spectrum should steepen as these additional sources fill out the shin part of the spectrum. (An energy spectrum will exhibit progressively heavier nuclei as observed.) Incident cosmic rays with rigidity well below the UHECR range will be convected adiabatically into the shock and see it as an essentially planar shock accelerator. Provided that the compression ratio of the subshock is neither too large, specifically $2.5\lesssim r\lesssim3.5$, then it will be the cosmic rays near the knee in the spectrum at few $PV$ that will dominate the UHECR spectrum that is reflected upstream back into the IGM. This will be reflected in the UHECR composition. In this model, the GeV particles that dominate the GCR spectrum and freshly injected particles are minor contributors to the output UHECR spectrum.     

How reasonable is it for a CAS to behave in this extreme fashion? And, by extension, how reasonable is it for SNR and GWTS to accelerate cosmic rays to energies $\sim 3$ and $\sim300\,{\rm PeV}$ respectively? Here, a rather different approach from that followed in most of the impressive simulations that have been completed may be helpful. Instead of starting from a quiescent initial state and monitoring the progressive acceleration of cosmic rays, and their associated local turbulence, to ever greater rigidity, limited by dynamic range and runtime, it might be interesting to start with the near-maximally turbulent state described and see if it can be self-sustaining. The dynamics is dominated by the highest energy particles and the background gas. PeV cosmic rays can be injected at the shock front. A key feature of this approach is that, as the rigidity increases, the particles diffuse further ahead of the shock front until they either escape upstream or are transmitted downstream with probabilities $\sim$ a half, ending their acceleration. It will also be necessary to include the curvature of the shock which facilitates this escape. These particles, with $R\sim R_{\rm max}$, will provide the first encounter that the inflowing intergalactic gas has with the shock. The gas itself is, presumably, still quite weakly magnetised with a pressure much smaller than the (anisotropic) pressure of the UHECR. Indeed, the pressure of the cosmic rays is likely to dominate the thermal pressure of the IGM. There is the possibility of hydromagnetic instability, similar to the firehose and mirror instabilities but also including the mean velocity of the UHECR fluid through the gas. This turbulence should grow as the gas falls further into the cluster with wavelengths longer than the local UHECR gyro radius ($\sim1-10B_{\mu{\rm G}}^{-1}\,{\rm kpc}$), but shorter than the distance from the shock. The turbulence, which should be quasi-stationary in space, will have decreasing outer scale as the shock is approached, and should evolve to shorter wavelengths, progressively scattering lower rigidity particles through strongly nonlinear, resonant interaction as the shock is approached. 

This ``bootstrap'' mechanism (adopting the physics as opposed to the statistics usage \citep{Blandford:2007zza}) provides a different approach to exploring the complex physics of DSA. In the context of cluster shocks, it needs to be augmented by the inclusion of photopion and photodisintegration loss which are important because, unlike with the SNS, GWTS cases, the acceleration timescale is not short compared with the propagation timescale. Indeed, intergalactic and Galactic diffusion needs to be included in the problem alongside the background velocity field which transitions from inflow to cosmological outflow about 7~Mpc away from the cluster for Virgo \citep{Shaya:2017wnh}. The complex thermal and multiphase nature of the IGM, which is likely to be influenced by the hypothesized magnetic turbulence, is also a large part of the problem. Finally, the downstream boundary condition on particle acceleration, which is normally idealized as strong scattering in a uniform flow could turn out to be important.

\section{Discussion and Future Goals}
Cosmic-rays have been known for almost one century and gave birth to particle physics, but we still don't know where they come from. The question of what else in the Galaxy besides supernovae makes cosmic rays  remains an open question and the origin of UHECR (cosmic rays with energy above 0.1 EeV) is still unknown, with both extragalactic and Galactic origin scenarios constrained but not ruled out. 

So far, only electromagnetic, gravitational, neutrino signals trace high energy sources. The possible candidate sources of UHECRs -- relativistic jets associated with binary neutron star mergers, core-collapse supernovae or active galactic nuclei; magnetars; accretion shocks around clusters of galaxies and filaments; wind bubbles associated with ultra fast outflows;  tidal disruption events -- still remain candidates and no direct evidence implicating them  has been found so far. The question of which  of these sources can be the accelerators of the ultra-high energy cosmic rays we detect up to   $\sim 200$ EeV still remains unanswered. Although much debated, the common picture is that a transition from a Galactic to an extragalactic origin for cosmic radiation occurs somewhere between the knee and the ankle, and that the cosmic-ray composition becomes heavier as the energy increases. The interpretation of the anisotropies that emerge at the highest energies needs a better understanding of the UHECR composition. 

The status is that the lack of observational understanding of the UHECR composition leaves freedom for  speculation about their origin. This is also an interesting problem in the understanding of the physics of the extensive air shower development; the observed muon number, which is is a key observable to infer the mass composition of UHECR, shows an excess compared to air shower simulations with state-of-the-art QCD models. This is known as the "Muon Puzzle" (\citep{Albrecht:2021cxw} for a recent account). Future data from the LHC, in particular oxygen beam collision would be needed to better address this issue. There is currently an effort from the UHECR community to better probe the cosmic ray composition  at the highest energies. AugerPrime, the ongoing upgrade of the PAO has been designed  to enhance the sensitivity of composition analyses by adding scintillator surface detectors and radio antennae to the existing  Cherenkov detectors \citep{Stasielak:2021hjm}. TA$\times$4, the ongoing upgrade of TA (an additional 500 scintillator surface detectors designed to provide a fourfold increase in effective area) is being deployed and is partly in operation \citep{bib:tax4_nim}. 

A few energy events have already been detected by Auger and TA above 150~EeV. With better statistics at the highest energies, it may be possible to unveil the sources \citep{Globus:2022qcr}.  The basic reason is that the background of distant sources, which dominates at lower energies, is less important at high energies as a consequence of the GZK cutoff.For example, at 300~EeV, only protons have a horizon $\gtrsim30$~Mpc, including  the Virgo cluster and M87 at $\sim$~16 Mpc. Any detection of intermediate or heavy mass nuclei at these energies would have to originate from sources at $\lesssim 3$~Mpc, again severely limiting the options for sources, probably pointing to transients in the local group.

The detection of doublets or multiplets of "extreme energy" cosmic ray events (i.e. UHECR with energy above 150 EeV) with a composition-sensitive detector could rule out many currently viable source models \citep{Globus:2022qcr}. Using recent Galactic magnetic field models \citep{Jansson:2012pc,TF17}, we calculated “treasure” sky maps to identify the most promising directions for detecting extreme energy cosmic-ray doublets, events that are close in arrival time and direction and predicted the incidence of doublets as a function of the nature of the source host galaxy. Based on the asymmetry in the distribution of time delays, we showed that observation of doublets might distinguish source models. This is again why larger exposures and a better approach to mass composition is needed.

Identifying the sources of UHECR, ushering in the dawn of cosmic ray astronomy and exploring the otherwise inaccessible fundamental physics involved constitute three strong reasons for taking cosmic-ray studies to the next level so as to measure individual cosmic-ray masses at the highest energy. Larger and more capable cosmic-ray detector arrays, such as the  Global Cosmic Ray Observatory \citep{horandel2022gcos} will be needed to accomplish this.

\bigskip
\section*{Acknowledgments}
We thank our colleagues that helped so much in shaping our understanding of UHECR physics and phenomenology: Denis Allard, Chen Ding, Glennys Farrar, Anatoli Fedynitch, Payel Mukhopadhyay, Etienne Parizot, Enrico Peretti, Paul Simeon, Alan Watson.
N.G.’s research is supported by the Simons Foundation, the Chancellor Fellowship at UCSC and the Vera Rubin Presidential Chair.\\

The breadth of topics briefly reviewed here is associated with an extensive bibliography which cannot be reflected in a proceedings. Apologies are proffered to colleagues whose important research is not cited.

\bibliography{scibib}
\end{document}